\documentclass[aps,prb,showpacs,superscriptaddress,twocolumn]{revtex4-1}
\usepackage{graphicx, epsfig, amssymb, amsmath, lineno, subfigure, float,lmodern}

\begin{document}
\pagenumbering{roman}
\title{Eddy-current effects on ferromagnetic resonance:  Spin wave excitations and microwave screening effects}

\author{Vegard Flovik}
\email{vflovik@gmail.com}
\affiliation{Department of Physics, Norwegian University of Science and
 Technology, N-7491 Trondheim, Norway}

\author{Bj{\o}rn Holst Pettersen}
\affiliation{Department of Physics, Norwegian University of Science and
  Technology, N-7491 Trondheim, Norway}

\author{Erik Wahlstr\"{o}m}
\affiliation{Department of Physics, Norwegian University of Science and
  Technology, N-7491 Trondheim, Norway}

\date{\today}

\begin{abstract}
We investigate how controlling induced eddy currents in thin film ferromagnet-normal metal (FM/NM) structures can be used to tailor the local microwave (MW) fields in ferromagnetic resonance (FMR) experiments. The MW fields produced by eddy currents will in general have a relative phase shift with respect to the applied MW field which depends on the sample geometry. The induced fields can thus partially compensate the applied MW field, effectively screening the FM in selected parts of the sample. 
The highly localized fields produced by eddy currents enable the excitation of spin wave modes with non-zero wave vectors ($k \neq 0$), in contrast to the uniform $k=0$ mode normally excited in FMR experiments. 
We find that the orientation of the applied MW field is one of the key parameters controlling the eddy-current effects. The induced currents are maximized when the applied MW field is oriented perpendicular to the sample plane. 
Increasing the magnitude of the eddy currents results in a stronger induced MW field, enabling a more effective screening of the applied MW field as well as an enhanced excitation of spin wave modes.
This investigation underlines that eddy currents can be used to control the magnitude and phase of the local MW fields in thin film structures.

\end{abstract}


\maketitle

\section{Introduction}

The microwave (MW) frequency spin dynamics in nanostructures usually involves stacks of layers combining ferromagnets (FM) and normal metals (NM) at the nanometer scale \cite{spindyn1,spindyn2}.
A time varying magnetic field with a component perpendicular to a conductive thin film induces circulating currents in the thin film plane. These currents, commonly referred to as eddy currents, produce secondary phase shifted magnetic fields in close proximity to the conductor.
The effects of eddy currents on ferromagnetic resonance (FMR) in conducting films are well known in the limit of film thickness approaching their electro-magnetic skin depth ( $\simeq$ 800 nm for bulk Au at 10 GHz). In those cases eddy-current effects can lead to FMR linewidth broadening, and spin-wave excitations due to inhomogeneous microwave fields \cite{eddy1,eddy2}. 

Eddy-current effects have often been neglected for film thicknesses below their skin depth. However, the contribution of the microwave conductivity of magnetic multilayers has received increasing attention in recent years, indicating the importance of eddy-current effects also for NM films far below their skin depth \cite{screening,screening2,screening_FMR, screening_FMR2,screening_FMR3,screening_FMR4,screening_FMR5,screening_FMR6,screening_FMR7,screening_FMR8,screening_FMR9,eddy_linewidth,eddy_linewidth2,eddy3}. 
In FM/NM bilayers Maksymov \textit{et. al.} showed that the amplitude of the magnetization precession in the FM layer can be diminished by the shielding effect due to microwave eddy currents circulating in the NM layer \cite{screening_FMR4}.
A study by Kostylev also showed that in single layer and bi-layer metallic FM the MW screening effect results in a spatially inhomogeneous MW field within the magnetic film \cite{screening_FMR9}. The experimental manifestation of this is a strong response of higher order standing spin wave modes due to the non-uniform MW field across the thickness of the magnetic film. 

In a recent study, Flovik \textit{et a}l. \cite{eddy3} investigated the effects of the induced Oersted fields in FM/NM bilayer structures. They show that the induced fields can strongly affect the FMR excitation, resulting in significant changes to the symmetry of the FMR lineshape. 
Differences in symmetry of FMR lines have been used to study the spin pumping from a magnetic material to a normal metal \cite{spinpump,spinpump2,spinpump3,spinpump4,spinpump5,spinpump6}. In such studies, lineshape symmetry is one of the main parameters used to analyze the results. Hence, to correctly interpret experimental data involving FMR it is important to understand how eddy currents affect the FMR excitation.

However, rather than considering eddy currents a parasitic effect, we here investigate how controlling the current paths can be used to tailor the local MW fields.
By using lithographically fabricated samples, we have systematically studied how sample and field geometry affects the coupling between the applied MW fields and eddy-current-induced fields. 
We show that eddy-current effects can have a significant impact on the FMR excitation even in very thin metallic FM ($\sim 10$ nm Py), determined by the sample and MW field geometry.
The induced MW fields from eddy currents can partially compensate the applied MW field, effectively screening the FM layer in selected parts of the sample.  
In contrast to the screening effects previously observed for continuous FM/NM bilayers by Maksymov \textit{et. al.}\cite{screening_FMR4}, we here consider the screening also in samples consisting of patterned NM structures.  

Controlling the current paths by patterned NM structures generates highly localized MW fields. We provide evidence that this enables the excitation of spin wave modes with non-zero wave vectors ($k \neq 0$), in contrast to the uniform $k=0$ mode normally excited in FMR experiments.
As we are considering very thin metallic FM ($\sim 10$ nm Py), we do not observe the aforementioned standing spin wave modes across the film thickness studied by Kostylev \cite{screening_FMR9}. 
Here, we argue that the inhomogeneous MW field produced by the induced eddy currents excite Damon-Eshbach surface spin wave modes\cite{DE_modes}.

The excitation of wave vector specific Damon-Eshbach spin waves in FM films using  a diffraction grating has been studied by Sklenar \textit{et. al} \cite{SW_excitation}. They showed that a patterned silver antidot lattice on a thin uniform permalloy film enables coupling to spin wave modes.
The amplitude of the spin wave excitations was however very small compared to the uniform FMR mode. 

Here, we show that the orientation of the applied MW field with respect to the sample plane is one of the key parameters controlling eddy-current effects. The induced currents are maximized when the applied MW field is oriented perpendicular to the sample plane.
Increasing the magnitude of the eddy currents results in a stronger induced MW field, enabling a more effective screening of the applied MW field as well as an enhanced excitation of spin wave modes.
This points towards the importance of considering eddy-current effects not only for understanding basic experiments, but also for control of the local MW field in thin film structures.

\section{Experimental setup and sample preparation}
\subsection{FMR experiments}
\begin{figure}[t]
\centering
\includegraphics[width=85 mm]{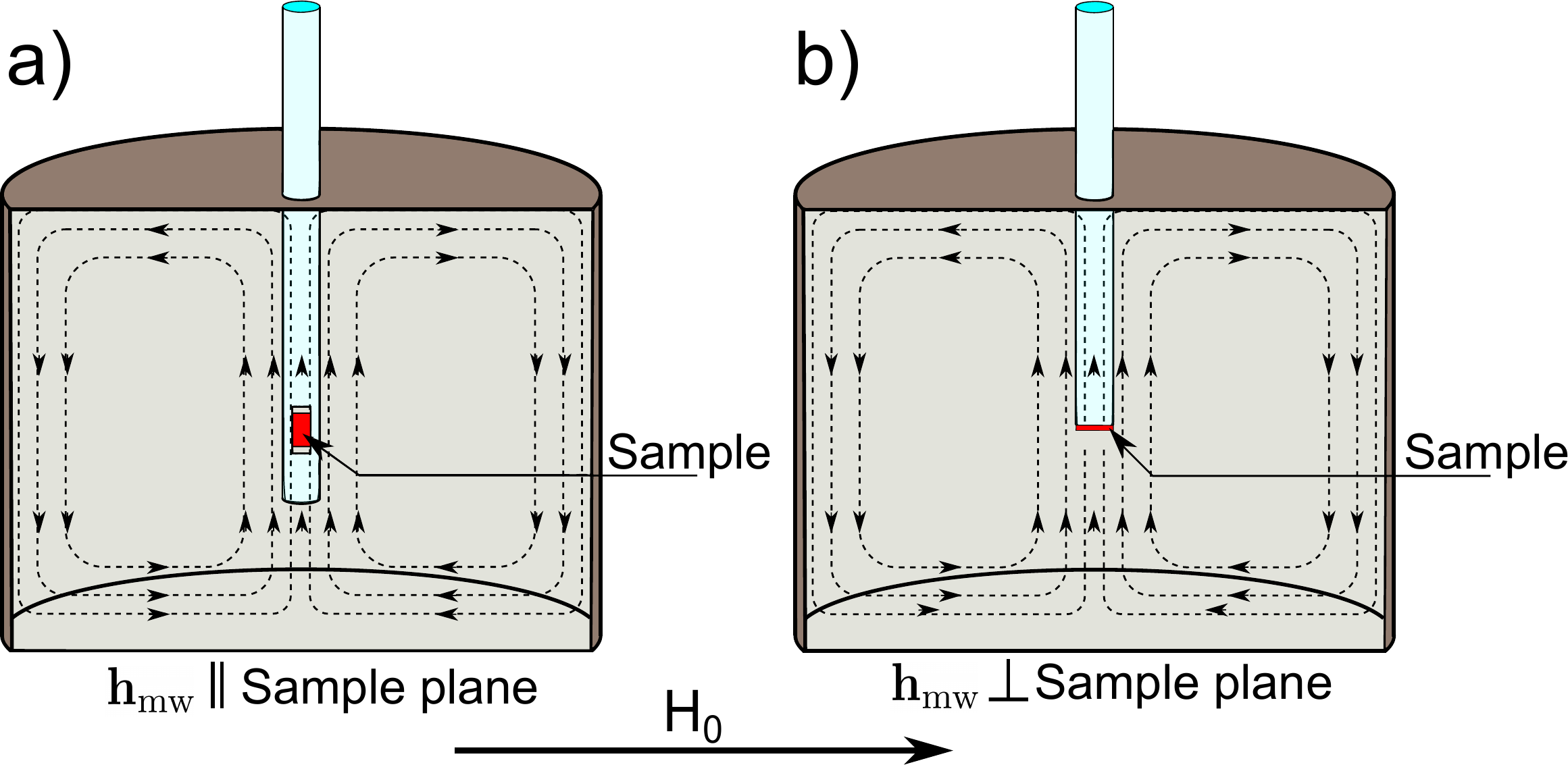}
\caption{\footnotesize Schematic of the cylindrical TE-011 microwave cavity, showing the sample position and field geometry for a) Vertical sample mounting resulting in $\bold{h}_{\text{mw}}$ parallel to the sample plane. b) Horizontal sample mounting resulting in $\bold{h}_{\text{mw}}$ perpendicular to the sample plane . A goniometer allows for a 360 degrees rotation of the sample with respect to the external static field, H$_0$. }
\label{fig:cavity}
\end{figure}

 Ferromagnetic resonance experiments were carried out in a commercial X-band electron paramagnetic resonance (EPR) setup with a fixed microwave frequency of 9.4 GHz (Bruker Bio-spin ELEXSYS 500, with a cylindrical TE-011 microwave cavity). The measurements were performed with a MW power of 0.65 mW, resulting in a MW field amplitude of $h_{\text{mw}} \approx  6 \mu$T at the sample position. The magnitude of the external field is then swept to locate the resonance field, $H_R$.
The sample is attached to a quartz rod connected to a goniometer, which allows for a 360 degrees rotation of the sample with respect to the external static field. The applied MW field is oriented either parallel (Fig. \ref{fig:cavity}a) or perpendicular (Fig. \ref{fig:cavity}b)  to the sample plane depending on sample mounting.
The MW field in the cavity can be considered uniform on the length scale of the sample, and rotationally symmetric due to the cylindrical shape of the cavity.

The ferromagnetic resonance is usually driven directly by the MW field from a cavity or from a coplanar waveguide/microstrip line. However, a time varying magnetic field component perpendicular to the sample plane will induce circulating eddy currents in conducting samples, and additional Oersted fields perturbing the FM. The FMR is often assumed to have a symmetric Lorentzian lineshape, but previous work by Flovik \textit{et al}. \cite{eddy3} show that eddy-current effects can strongly affect the lineshape symmetry.
We thus fit the FMR lineshapes to a linear combination of symmetric and antisymmetric contributions, determined by the $\beta$ parameter in Eq.(\ref{eq:fiteq}).

\begin{equation}\label{eq:fiteq}
\chi = A \frac{1+\beta(H_0-H_R)/\Gamma}{(H_0-H_R)^2 + (\Gamma /2)^2}.
\end{equation}

Here, $A$ is a general amplitude prefactor, $H_0$ and $H_R$ are the external field and resonance field respectively, and $\Gamma$ the FMR full linewidth at half maximum, FWHM.  
This expression consists of two components: a symmetric absorption lineshape arising from the in-phase driving fields, and an antisymmetric dispersive lineshape proportional to $\beta$ arising from out-of-phase driving fields from the induced eddy currents. In this form, Eq.(\ref{eq:fiteq}) describes what is known in the literature as Dysonian lineshapes \cite{dyson, oates, poole}.

The FMR experiments were performed with a low amplitude ac modulation of the static field, which allows lock-in detection to be used in order to increase the signal to noise ratio. The measured FMR signal is then proportional to the field derivative of the absorption, and the experimental data was thus fitted to Eq.(\ref{eq:fiteq2}), $d\chi/ d H_0$

\begin{multline}\label{eq:fiteq2}
\frac{d\chi}{dH_0} = A \Bigg [\frac{\beta/ \Gamma }{(H_0-H_R)^2 + (\Gamma /2)^2} \\
 -  \frac{2(H_0-H_R)[1+\beta(H_0-H_R)/ \Gamma]}{[(H_0-H_R)^2 + (\Gamma /2)^2]^2} \Bigg ].
\end{multline}
\\

\subsection{Sample preparation}\label{sec:sampleprep}

Experiments were performed with Permalloy (Py=Fe$_{20}$Ni$_{80}$) as the FM layer, and gold (Au) as the NM layer. The thin film structures were prepared using a lift off process combining optical lithography, DC magnetron sputter deposition and electron beam evaporation. 

A thin layer of AZ5214E image reversal resist was first applied to a clean silicon support. To enhance the resolution of the lithographic pattern transfer, edge beads corresponding to resist thickness variations along the
sample edge are removed before transferring the pattern from a photomask to the resist layer. Before loading the resist covered substrates into the sputtering chamber a hard baking step was performed to reduce water content in the photoresist, thereby increasing its etch resistance. After loading the patterned substrate into the vacuum chamber an argon pre-sputtering step was performed to remove contaminants from the sample surface, thus improving the quality of the final thin film. Permalloy was deposited using DC magnetron sputtering. The steps following Py deposition depended on the number of lithography steps used to define the final thin film structure. In the single mask process, gold was deposited directly on top of Py by E-beam evaporation. A lift off process was performed to dissolve the photoresist, leaving only the thin film stacks in direct contact with the substrate. In the multi mask process, lift off was performed directly after Py deposition followed by a second lithography process transferring a secondary pattern to the substrate, thus allowing one to vary the geometry of the gold layer independently of the Py layer. After completing the final lift off process all samples were inspected by optical microscopy to reveal defective samples.  Atomic force microscopy (AFM) was used to measure the height of deposited thin films, and the individual samples produced on the same silicon support were separated using an automated scriber and breaker.

\section{Theoretical considerations}

\subsection{Eddy current induction in a NM thin film disc}

\begin{figure}[t]
\centering
\includegraphics[width=85mm]{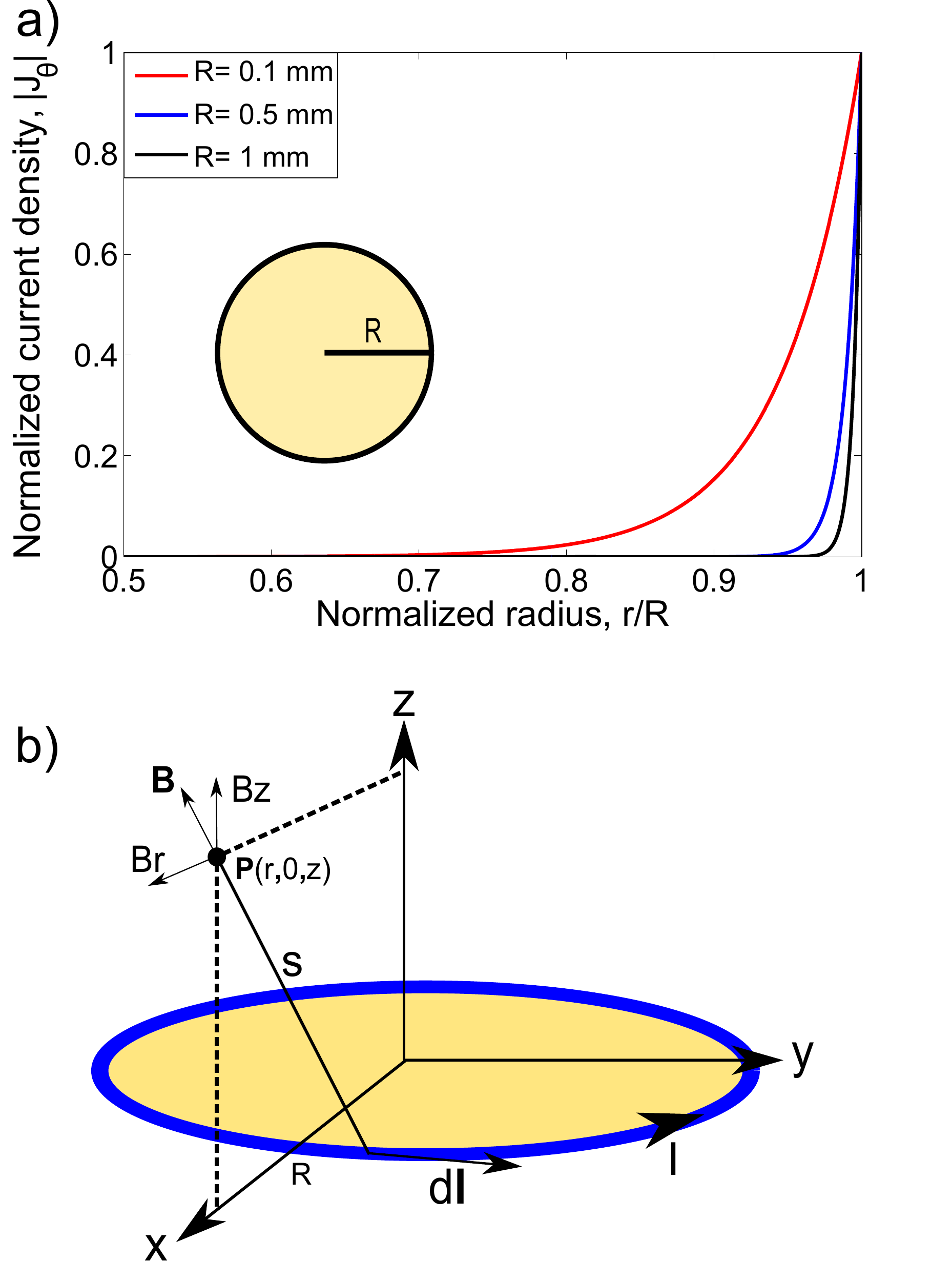}
\caption{\footnotesize a) Calculated current density from Eq.(\ref{eq:CurrentDensity}) as a function of $r$ for samples of radius R. b) Geometry of the circular current loop}
\label{fig:current_distr}
\end{figure}

A numerical approach is generally needed to calculate the distribution of induced currents and the associated magnetic fields. However, a closed form solution is obtainable for the simple case of a circular metallic film of thickness much less than the electro-magnetic skin depth. 
For a spatially homogeneous time harmonic magnetic field \(\mathbf{B}e^{i\omega t}\) applied perpendicular to a non-magnetic circular disc, 
the induced current density in the thin film plane can be described by\cite{ind_eddycurr}

\begin{equation}
\label{eq:CurrentDensity}
J_{\phi}(r)=-\frac{k|\mathbf{B}|}{\mu_0}\frac{I_1(kr)}{I_0(kR)},
\end{equation}
where 
\begin{equation}
k=\sqrt{\omega\mu_0\sigma}e^{i\pi/4}.
\end{equation}
\noindent
Here, $R$ is the radius of the disc, \(\mu_0\) the vacuum permeability, \(\sigma\) the film conductivity and \(I_n(\alpha)\) the modified Bessel function of the first kind and order n. In Fig. \ref{fig:current_distr}a we plot the normalized current density calculated from Eq.(\ref{eq:CurrentDensity}) along a circular sample for various radii in the range r=[0.1,1] mm, for a conductivity of $\sigma_{\text{py}}\approx 3 \cdot 10^{6}$ S/m \cite{cond_Py} and a microwave frequency of 9.4 GHz. 

As shown in Fig. \ref{fig:current_distr}a, the current density is localized primarily along the the sample edge as the disc size is increased. 
The current distribution also depends strongly on MW frequency, with a narrower distribution as the frequency is increased. However, in this work we consider a fixed microwave frequency of 9.4 GHz. Due to the localized current distribution, we approximate to first order the induced current as a single circular current loop along the sample edge. 

\subsection{Magnetic field from a circular current loop}\label{sec:currentloop}

For a circular loop carrying a current I, the magnetic field at any point in space can be obtained from the magnetic vector potential:

\begin{equation}
\label{eq:Vector potential loop}
\mathbf{A}=\frac{\mu_0 I}{4\pi}\oint \frac{d\mathbf{l}}{s},
\end{equation}
\noindent
s here being the distance from a point in space, P, to the line element \(d\mathbf{l}\), as shown in Fig. \ref{fig:current_distr}b. 
The general solution to Eq.(\ref{eq:Vector potential loop}) yields the vector potential \cite{jackson}:

\begin{equation}
\mathbf{A}=A_\phi\boldsymbol{\hat{\phi}}=\frac{\mu_0 I}{2\pi}[2k^{-1}r^{-/2}(K(k)-E(k))-kr^{-1/2}K(k)]\boldsymbol{\hat{\phi}}.
\end{equation}
\noindent
Here \(K\) and \(E\) represent complete elliptical integrals of the first and second kind respectively, while 
\begin{equation}
k=\sqrt{\frac{4rR}{z^2+(R+r)^2}}.
\end{equation}

\noindent
From the vector potential $\mathbf{A}$, one can calculate the magnetic field (\(\mathbf{B}=\nabla\times\mathbf{A}\)): 
\begin{align}
\label{eq:Bz}
B_z(r,z) & = \frac{\mu_0 I}{2\pi\sqrt{z^2+(R+r)^2}}\left(\frac{R^2 - z^2 - r^2}{z^2 + (r-R)^2}E(k)+K(k)\right) 		\\\nonumber\\
\label{eq:Br}
B_r(r,z) & = \frac{\mu_0 I z}{2 \pi r \sqrt{z^2+(R+r)^2}}\left(\frac{R^2 + z^2 + r^2}{z^2 + (r-R)^2}E(k)-K(k)\right)
\end{align}

In previous work by  Flovik \textit{et al.} \cite{eddy3}, it was shown that phase shifted contributions to the FMR excitation produced by eddy currents results in an asymmetry of the observed lineshapes. Other studies have also shown that the relative phases of electromagnetic waves are important to consider in FMR exeriments \cite{review_FMR,lineshape1}. 
Hence, the phase lag between the applied MW field and the induced field needs to be considered.  
The Oersted fields produced by the eddy currents have a relative phase lag, $\phi$, compared to the applied MW field, which in the ideal case is expected to be $\phi=-90$ degrees ($I_\text{Eddy} \propto \frac{\partial h}{\partial t}$). 
However, due to the inductance and resistance of the film, there will be an additional phase between the applied MW field and the induced field. 
At larger film thicknesses, one also needs to take into account phase shifts due to the skin effect. A complex system such as an experimental setup involving waveguides, coaxial cables etc. can also introduce a non-zero phase offset, $\phi_0$ \cite{review_FMR,lineshape1}. Considering this, one can write the relative phase lag as \cite{jackson}:  

\begin{equation}\label{eq:phase}
\phi= -\left[90  + \tan^{-1} \left( \frac{\omega L}{R} \right) + d/\delta + \phi_0 \right].
\end{equation}
\begin{figure*}[]
\centering
\includegraphics[width=170 mm]{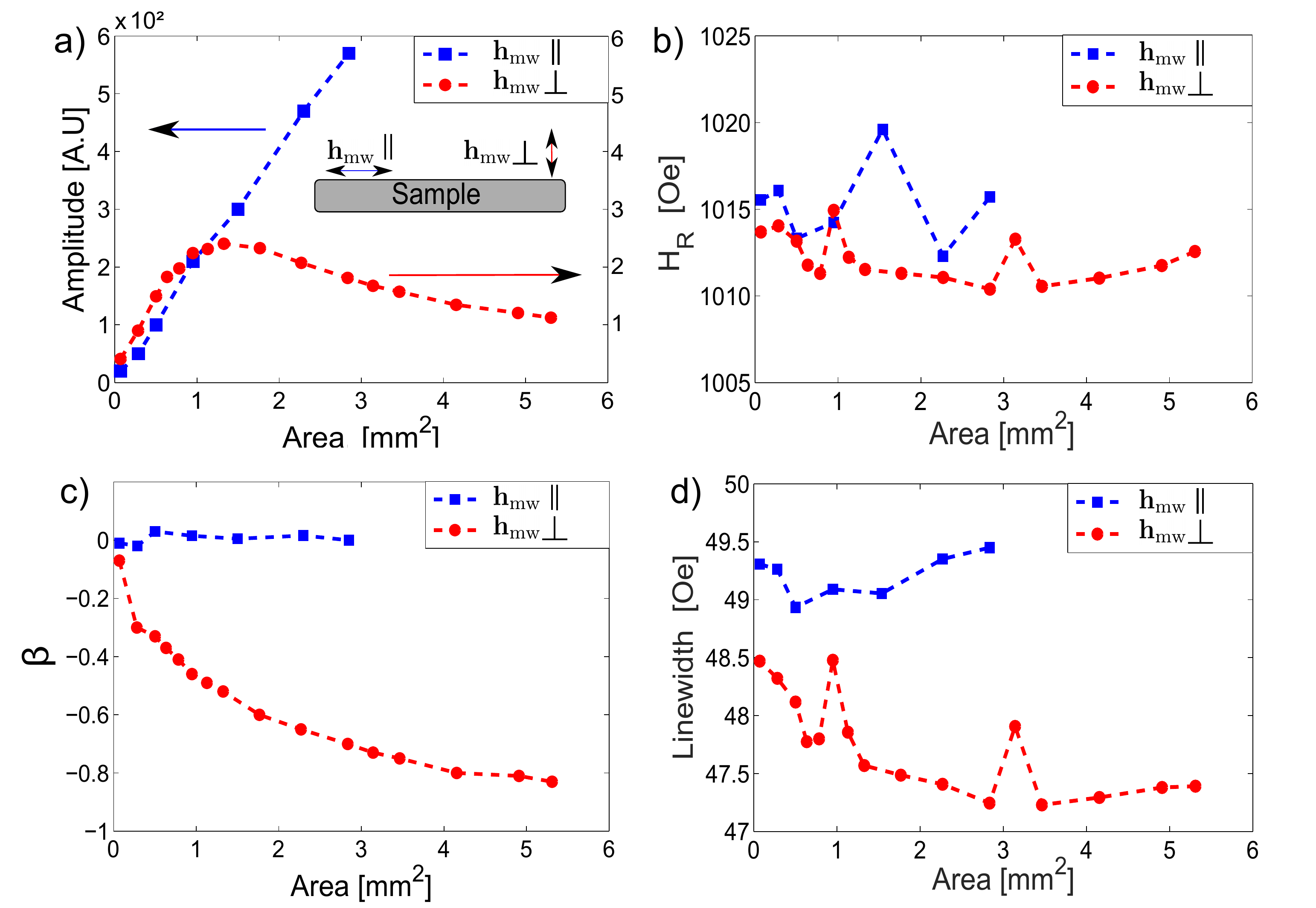}
\caption{\footnotesize a)  FMR absorption amplitude  as a function of disc area for the MW field oriented parallel and perpendicular to the sample plane respectively. Inset: Geometry for MW field oriented parallel and perpendicular to the sample plane. b) Resonance field, c) Lineshape asymmetry parameter $\beta$ and d) Linewidth for the same field geometries.  }
\label{fig:amplitude2}
\end{figure*}
\noindent
Here, $\omega$ is the microwave angular frequency, $L$ and $R$ are the inductance and resistance of the film, $d$ is the film thickness and $\delta$ is the MW skin depth ($\simeq 800$ nm for Au at 10 GHz).

\section{Results and discussion}
The effects of eddy currents on the FMR excitation depends on both sample and field geometry. In sec. \ref{sec:Py_single} we investigate the effects of the MW field geometry, and show that eddy-current effects are significant also for thin FM layers ($\sim$ 10 nm Py). In sec. \ref{sec:PyAu} we add NM layers with a high conductivity compared to the Py layer.  The induced currents in the NM layer produce localized MW fields, enabling the excitation of spin wave modes with a non-zero wavevector ($k \neq 0$). In sec. \ref{sec:PyAu_ring} we investigate how patterned NM structures can be used to control the induced current paths, and by this the local MW field in thin film structures. 

\subsection{Circular Py discs: Effect of MW field geometry}\label{sec:Py_single}

We first characterize the simplest case of samples consisting of a single layer of Py, and investigate the FMR absorption in a series of Py discs with a thickness of 10 nm and radius in the range 0.15-1.3 mm. 

By fitting the FMR lineshape to Eq.(\ref{eq:fiteq2}), we plot the absorption amplitude, resonance field, lineshape assymetry parameter $\beta$ and linewidth as a function of disc area in Fig. \ref{fig:amplitude2}. Results are shown for the applied MW field oriented both parallel and perpendicular to the sample plane. 
Having the MW field oriented perpendicular to the sample plane will according to Faraday`s law maximize the induced currents in the sample, while rotating the MW field parallel to the sample plane should minimize any eddy-current effects. 
The FMR absorption amplitude is proportional to the energy dissipation in the sample. For the MW field oriented in the sample plane such that eddy-current effects are minimized, we observe the expected behavior that the FMR absorption amplitude increases linearly with sample volume (blue squares in Fig. \ref{fig:amplitude2}a). The lineshape also remains symmetric for all sample sizes with an asymmetry parameter $\beta \approx 0$ for the field geometry that minimize eddy-current effects (blue squares in Fig. \ref{fig:amplitude2}c).
The lineshape asymmetry in the geometry that maximize eddy currents  (red dots in Fig. \ref{fig:amplitude2}c) is due to the contribution of phase shifted induced fields from eddy currents to the FMR excitation, as discussed in sec. \ref{sec:currentloop} and in a previous study \cite{eddy3}. 

The differences in resonance field and linewidth  (Fig. \ref{fig:amplitude2}b and d) between the two geometries can be explained by effects which we do not attribute to eddy currents. Due to different sample holders for the two geometries there could be a slight misalignment of the external static field with respect to the sample plane, resulting in a small shift in resonance field and linewidth. Another consideration is that the magnetization precession in thin films  is not circular but elliptic due to strong demagnetizing fields, which force the magnetization into the film-plane. Having the applied MW field oriented along the long/short elliptic trajectory in the two geometries will affect how effectively the FMR is excited. This determines the cone angle of the magnetization precession, which could also cause a small shift in resonance field and linewidth. 

The significant difference between the two MW field geometries is thus the FMR absorption amplitude and lineshape asymmetry. 
Comparing the results, one notice that for small samples the amplitude increases with sample size in both cases, until the sample reaches an area of approximately 1 mm$^2$ (Fig. \ref{fig:amplitude2}a). Above this size, the amplitude starts to decrease as the sample size is increased when the MW field is oriented perpendicular to the sample plane. The lineshape asymmetry $\beta$ also increases with sample size, reaching a limiting value of $|\beta| <1$ for large samples (Fig. \ref{fig:amplitude2}c).  

While these trends present strong evidence for the effects of eddy currents on FMR in a single layer of Py, care should be taken when comparing the absorption amplitude for the two field geometries. As shown in Fig. \ref{fig:amplitude2}a, an in-plane orientation of the MW field results in an increase in amplitude compared to the same samples measured with an out-of-plane MW field. This is partially attributed to the screening effect of the applied MW field by the induced fields from eddy currents, but there are additional factors which also needs to be considered. In the two measurement geometries the sample orientation with respect to the magnetic mode in the cavity is different, which could affect the detector coupling coefficient $\kappa$. As the recorded signal amplitude is proportional to $\kappa$, this would introduce a scaling offset in the signal amplitudes. 
Another consideration is the aforementioned ellipticity of the magnetization precession in thin films. Having the MW field oriented along the long/short elliptic trajectory in the two geometries will affect the FMR excitation, and by this the absorption amplitude. 
The difference in amplitudes between the two geometries could thus be caused by a combination of several factors. 
The main point to note is rather the dependence of absorption amplitude vs. sample area, whether a linear trend is observed or not.
To explain the observed trend in the FMR absorption amplitude vs. disc radius in Fig. \ref{fig:amplitude2}a, which show a maximum for an intermediate disc radius, we consider a simple current loop model: 

From electromagnetic theory we know that the induced electromotive force (EMF), $\epsilon$, is proportional to the time derivative of the magnetic flux enclosed by the sample, and $\epsilon$ will thus increase proportional to the sample area. At the same time, approximating the current path as a circular loop around the edges of the sample, the resistance, R,  of the loop scales linearly with the radius $r$. This means that the magnitude of the induced current should scale linearly with the sample radius, as $|I_{\text{ind}}|=|\epsilon| / R \propto \pi r^2 /2 \pi r$.  
As the simplest case we consider the magnetic field from a circular current loop, given by Eqs.(\ref{eq:Bz}) and (\ref{eq:Br}). For a single layer of Py, we are interested in the magnetic field in the sample plane. Calculating the magnetic field at the center of the current loop in the x-y plane, the expressions simplify to the well known result for the magnetic field from a circular current loop: $B_z=\mu_0 I_{\text{Ind}}/2 \pi r$. As mentioned previously, the magnitude of the induced current should scale linearly with the disk radius. Using these approximations results in that the magnetic field at the center of the loop is independent of the loop radius. 
This can not simply explain the observed maximum in the FMR absorption amplitude for an intermediate disc radius in Fig. \ref{fig:amplitude2}a.
However, one of the approximations used was that one could consider the induced current as a single current loop localized at the sample edge. As shown in Fig. \ref{fig:current_distr}a, this approximation is not independent of the disc radius. For small discs, even if the current density is highest along the edges, it is still more evenly distributed throughout the sample. The current distribution also determines the balance between in-plane and out-of-plane components of the corresponding induced MW field perturbing the FM. 

Determining the relative magnitude of the MW field components is in general a complicated problem, and would require a simultaneous numerical solution of the coupled Maxwells and LLG equations \cite{LLG} for the various sample geometries. However, previous observations indicate that the induced field perturbing the FM can be comparable and in some cases even larger than the applied MW field, due to the close proximity to the induced currents at the FM/NM interface \cite{eddy3}. 
Two limiting cases are worth considering: A uniform current distribution in the sample would result in induced field components having only in-plane components (analogous to the magnetic field from a current in an infinite conducting plane). On the other hand, a localized current at the sample edge would produce mainly out of plane components of the corresponding magnetic field in the sample plane. As the applied MW field is oriented perpendicular to the sample plane, having induced field components perpendicular to the plane is needed in order to partially compensate the applied field. This could thus explain the observed behavior in Fig. \ref{fig:amplitude2}a: For the smallest samples, having a more uniform current distribution, the induced field components will have significant components in the sample plane. As the sample size is increased and the current density more localized along the sample edge, the induced fields will have the main component perpendicular to the sample plane, and could thus partially compensate the applied MW field. A compensation effect, reducing the effective MW field exciting the FMR, would lead to the decrease in the absorption amplitude which is observed for the larger samples. 
\newpage

\subsection{Circular Py discs with Au capping: Spin wave excitations}\label{sec:PyAu}
\begin{figure}[b]
\centering
\includegraphics[width=90 mm]{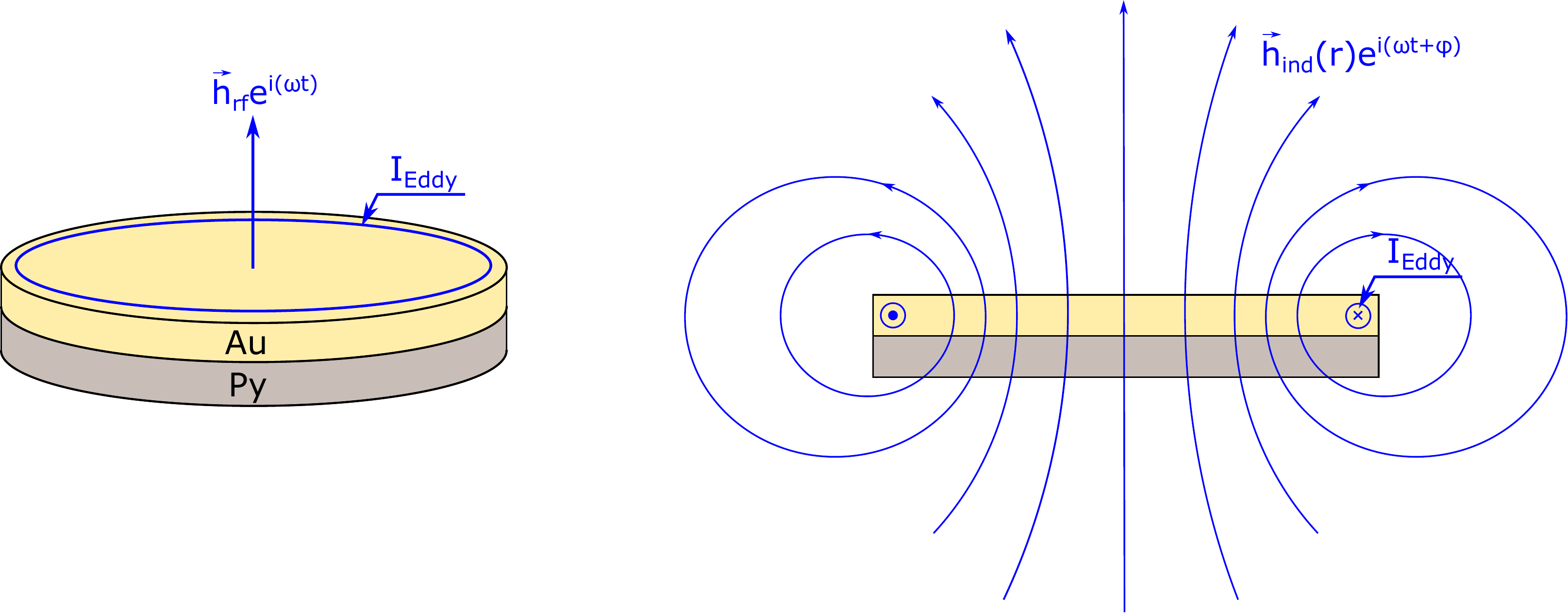}
\caption{\footnotesize Sample and field geometry, Py/Au bilayers  }
\label{fig:field_geometry}
\end{figure}

By adding NM layers with a high conductivity compared to the Py layer one can control the dominant current paths, and thus also the induced MW fields.
We fabricated two separate set of samples consisting of Py discs capped with a 10 nm Au layer, as shown in Fig. \ref{fig:field_geometry}. The conductivity of Au is significantly higher than for Py, with $\sigma_{Au}/\sigma_{Py} \approx 6-7$ \cite{cond_Py}. This results in larger induced currents in the sample, mainly flowing in the Au layer. 

\begin{figure*}[]
\centering
\includegraphics[width=170 mm]{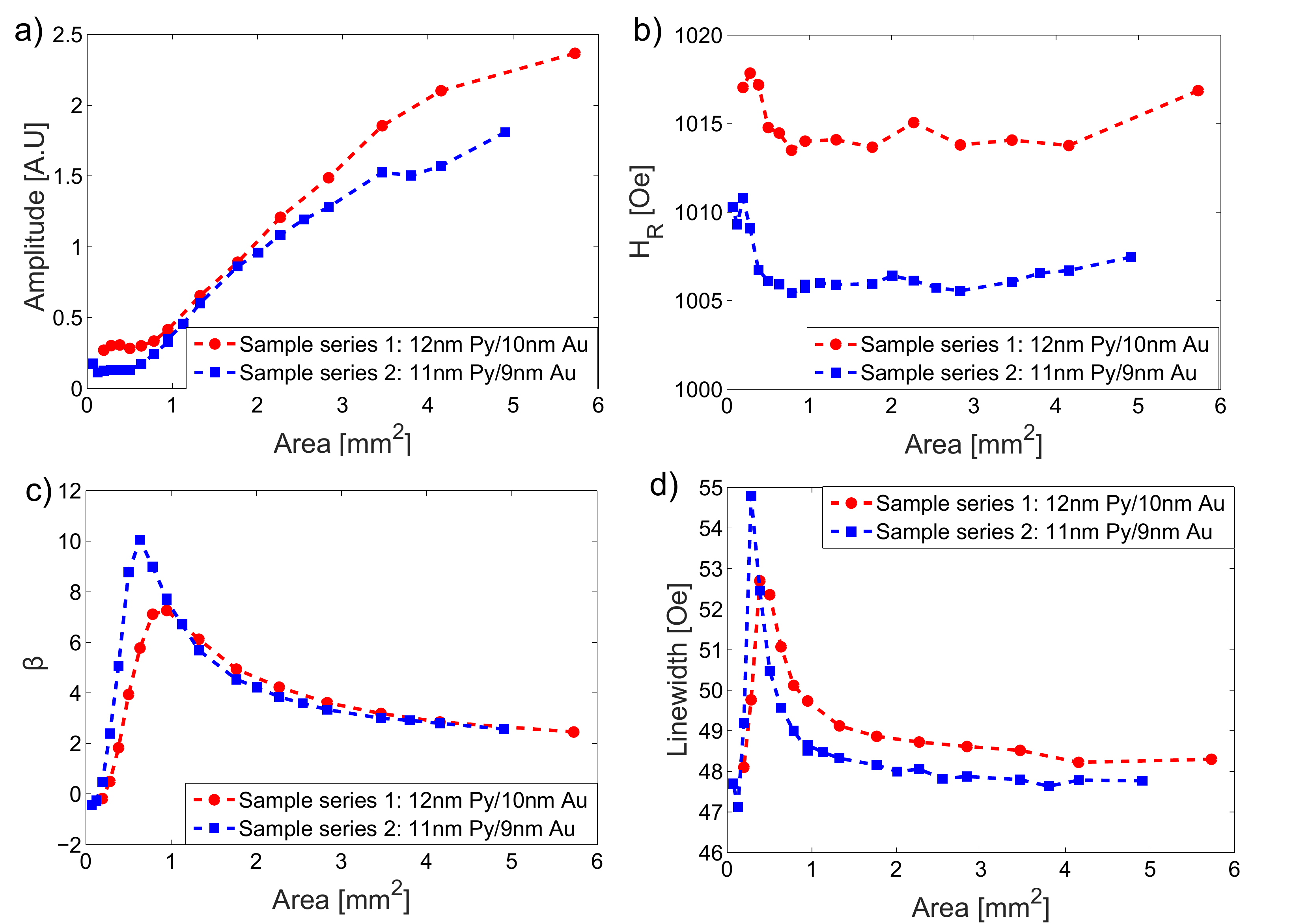}
\caption{\footnotesize  Comparing two set of samples of Py capped with a thin Au layer. a)  FMR absorption amplitude, b) Resonance field, c) Lineshape asymmetry parameter $\beta$, d) Linewidth  }
\label{fig:PyAu_all}
\end{figure*}

The two set of samples both give similar results. Starting with the absorption amplitude in  Fig. \ref{fig:PyAu_all}a, it does not follow the same trend as for the single layer Py samples (Fig. \ref{fig:amplitude2}a). For a disc area below $0.5 \text{mm}^2$, there is an initial region where the signal amplitude increases slowly with sample size, whereas for larger samples a clear increase in amplitude with sample size is observed. 
This initial region corresponds with the size range where a peak in both lineshape asymmetry $\beta$ and linewidth is observed (Fig. \ref{fig:PyAu_all}c and d). There is also a shift in the resonance field of $\approx$ 5 Oe, before remaining relatively constant for larger samples (Fig. \ref{fig:PyAu_all}b). The shift in resonance field and linewidth between the two sample series can be explained by a slight variation in material properties, as the samples were deposited at two separate occasions. 

Compared to the case of a single layer of Py, there are some important factors that differ: With a circulating current flowing in the Au layer, the Py layer will experience MW field components in both the in-plane and out-of-plane orientation, as indicated in Fig. \ref{fig:field_geometry}. 
The balance between in-plane vs. out-of-plane components is determined by the current distribution in the sample, as regions directly below the current paths will be dominated by in-plane components. As the sample size is increased, the current paths will be gradually more localized along the sample edge (Fig. \ref{fig:current_distr}a).  As this occurs, the induced field will gradually be dominated by out-of-plane components (Fig. \ref{fig:field_geometry}). This means that one should expect three different size regimes governing the influence by eddy-current effects.

%\newpage

For the smallest samples the induced currents are low due to the small size of the sample, as well as having a more uniformly distributed current density (Fig. \ref{fig:current_distr}a). Still, there are observable eddy-current effects. 
For a disc area below $ \approx 0.5 \text{mm}^2$, there is an initial region where the signal amplitude increases slowly with sample size compared to larger samples (Fig. \ref{fig:PyAu_all}a). This behavoir is attributed to screening effects, reducing the amplitude of the magnetization precession in the FM layer. 

For increasing sample sizes the current density becomes gradually more localized along the sample edges (Fig. \ref{fig:current_distr}a). The FM will experience both in-plane and out-of-plane MW field components, and this regime corresponds to where we observe significant changes in behavior for both amplitude, resonance field, lineshape asymmetry and linewidth (Fig. \ref{fig:PyAu_all}a-d). 
It has been shown that non-uniform MW fields can excite spin waves with a non-zero wave vector ($k\neq0$), and that this can cause a shift in resonance frequency and broadening of the FMR linewidth \cite{eddy_linewidth,eddy_linewidth2}. Here, the region which has dominant contributions from both in-plane and out-of-plane MW field components, and thus a non-uniform  MW field, is localized along the sample edge. The width of this edge region will thus determine the range of possible wave vectors of spin wave excitations. The excitation of Damon-Eshbach surface spin wave modes is dominating in this field geometry \cite{DE_modes, sw_excitation}, and the spin wave frequency is determined by the magnetostatic spin wave dispersion given by the following \cite{sw_disp}:

\begin{equation}\label{eq:wk}
w(k)=\gamma \mu_0  \left[H_0 (H_0 + M_s) + M_s ^2 \frac{ (1 - e^{-2 k d} )}{4} \right]^{1/2},   
\end{equation}
\noindent
where $\gamma$ is the gyromagnetic ratio, $\mu_0$ the vacuum permeability, $H_0$ the external field, $M_s$ the saturation magnetization of Py, $k$ the spin wave vector and $d$ the thickness of the Py film. 

\begin{figure}[t]
\centering
\includegraphics[width=90 mm]{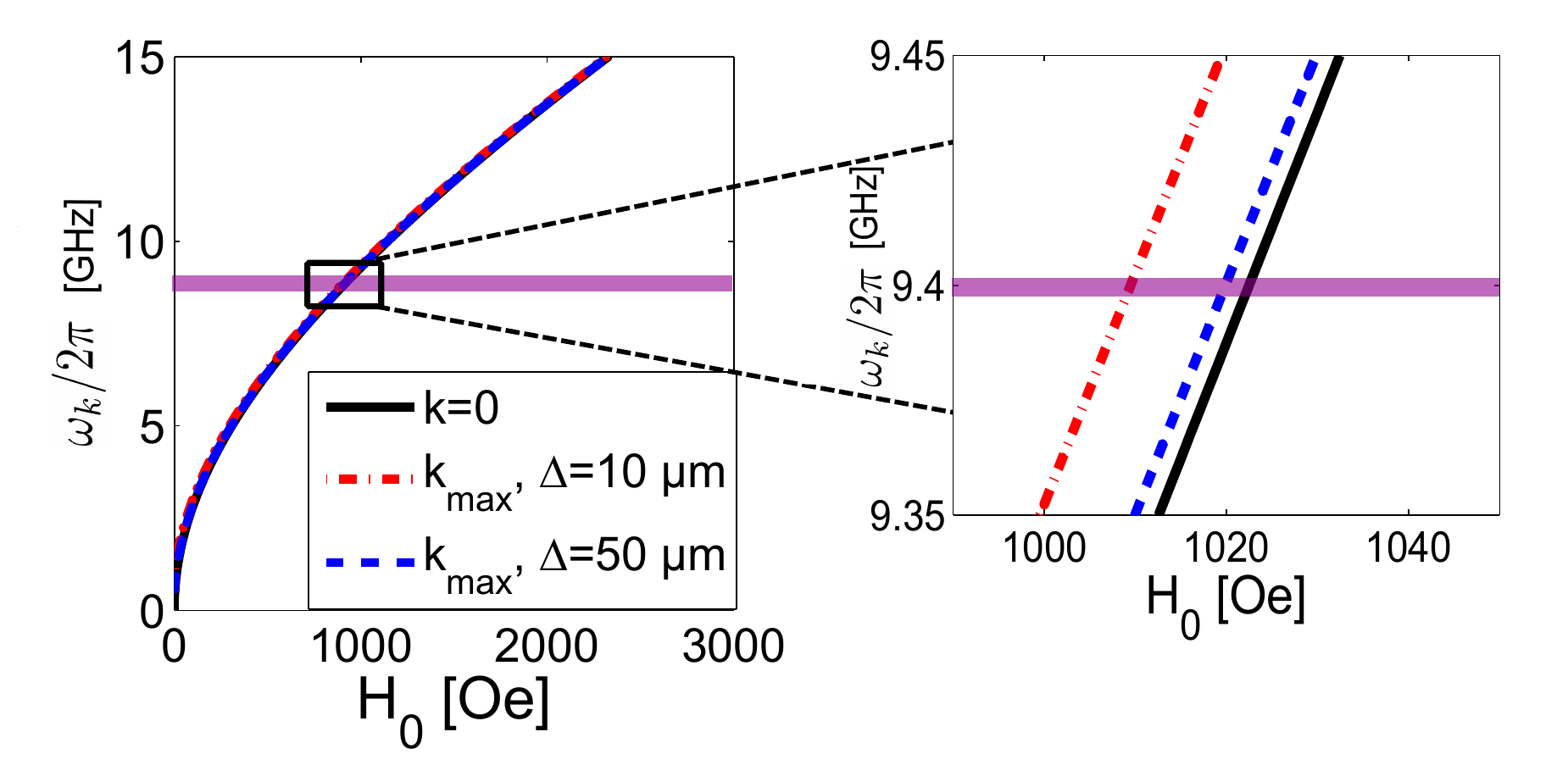}
\caption{\footnotesize Calculated spin wave dispersion, $\omega (k)$ from Eq. (\ref{eq:wk}) for spin waves with wave vectors $k_{\textit{max}}$ , determined by the width of the edge region, $\Delta$. The cavity FMR frequency of 9.4 GHz is shown as solid line.  }
\label{fig:wk}
\end{figure}

If we denote the width of the edge region by $\Delta$, the maximum spin wave vector will be determined by $k_{\text{max}}\propto \pi / \Delta$. 
From the radial current density in Fig. \ref{fig:current_distr}a, the area with highest current density is localized in an edge region of width $\Delta$ $\approx$ 10-50 $\mu$m.
As a simple estimate we calculate the spin wave dispersion from Eq.(\ref{eq:wk}) for the uniform $k=0$ mode as well as the spin wave modes for $k_{\text{max}}$ when $\Delta=10$ $\mu$m and $\Delta=50$ $\mu$m. The results are shown in in Fig. \ref{fig:wk}, where the FMR cavity resonance frequency is indicated by the solid line at 9.4 GHz. Shown as inset is a zoom-in of the same plot, where the splitting between the various modes is visible. The calculated splitting between the $k=0$ mode and the spin wave modes is 3 and 13 Oe respectively, determined by the width of the edge region, for $\Delta=50$ $\mu$m and $\Delta=10$ $\mu$m. 

The excitable spin wave modes vary over a continuous range, $0 < k <k_{\text{max}}$. In the linear response regime, the total FMR response can be taken as the sum of all excitable states. As shown in the inset of Fig. \ref{fig:wk}, for such long wavelength spin waves the shift in resonance field between excitations in the range $0< k <k_{\text{max}}$ is to small to be observed as separate peaks in the FMR spectrum. Summing the FMR response over a continuum of closely spaced states would thus rather appear as an apparent broadening of the FMR linewidth, where the broadening will be in the order of the shift in resonance field between the modes. 
This suggests that the observed linewidth broadening of $\approx$ 5 Oe in Fig. \ref{fig:PyAu_all}d for intermediate sample sizes is related to the excitation of spin waves modes with $k \neq 0$, due to the mixing of in-plane and out-of-plane MW field components for these sample sizes. By further increasing the sample size the FM will experience mainly out-of-plane field components, and one would no longer expect significant contributions from spin wave excitations to the total FMR absorption. This corresponds well with the experimental observations, where the resonance field and linewidth approach constant values for large samples.

For the lineshape asymmetry in Fig. \ref{fig:PyAu_all}c we observe a similar trend as for the linewidth, with a maxima for intermediate sample sizes.  
As mentioned previously, the lineshape asymmetry results from several phase shifted contributions to the FMR excitation. One would thus expect a maxima in $\beta$ in the size range where the FM experience a mixing of several MW field components, which is consistent with the experimental observations. 
The asymmetry parameter can have both positive and negative sign, depending on sample geometry and the phase shifts between the applied MW field and the induced fields. 
In this geometry, we observe a sign change in $\beta$ compared to the case of single Py discs in sec \ref{sec:Py_single}. 
This is caused by the change in sample geometry where the current in this case is mainly flowing in the adjacent NM layer, compared to in the FM layer for the first samples.
The different sample geometry introduces an effective phase shift to the induced field perturbing the FM, and thus a sign change in $\beta$. 
In general, determining the magnitudes and phases of the MW fields exciting the FMR is a non-trivial problem to solve even with a numerical approach.
However, the lineshape asymmetry is not our main focus, and we refer to earlier work investigating the effects of eddy currents on the FMR lineshape \cite{eddy3}. 

Our results indicate that that by modifying the conductivity one can control the dominant current paths in the sample, and thus the induced MW fields. As a model system to investigate this further, we fabricated patterned NM structures which enable better control of the current paths.

\subsection{Circular Py discs enclosed by Au ring: MW screening effects}\label{sec:PyAu_ring}

\begin{figure}[b]
\centering
\includegraphics[width=90 mm]{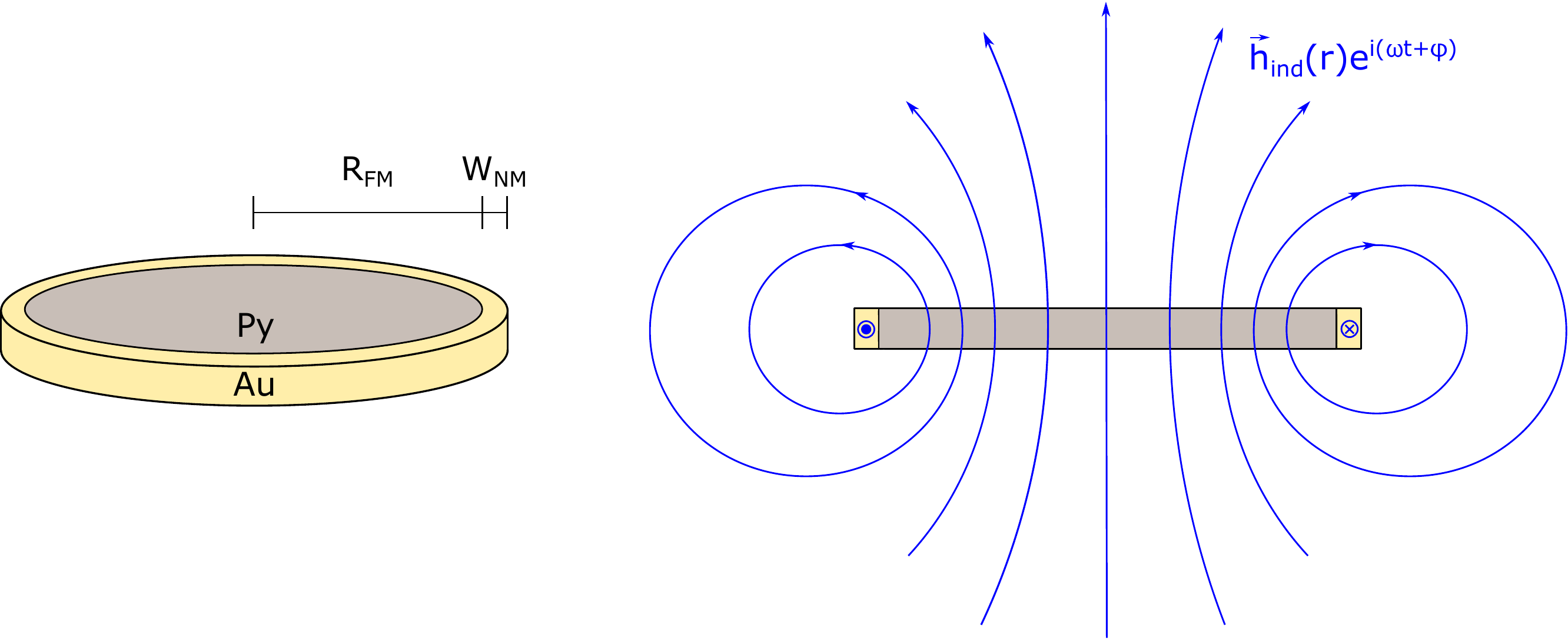}
\caption{\footnotesize Sample and field geometry Py/Au ring series  }
\label{fig:field_geometry_halo}
\end{figure}

As a final set of samples, we fabricated structures consisting of Py discs enclosed by an Au ring, as shown in Fig. \ref{fig:field_geometry_halo}. 
The high conductivity of Au compared to Py, as well as the radial current density distribution shown in Fig. \ref{fig:current_distr}a, results in that the induced current is mainly flowing in the Au ring. In this case the induced MW field perturbing the FM will have only out-of-plane components, as the radial component vanish due to the sample geometry. 
The fabricated samples consists of series where both the radius of the FM disk, as well as the width of the NM ring is varied. Increasing the NM ring width reduces its resistance, resulting in stronger eddy-current effects. 

\begin{figure*}[]
\centering
\includegraphics[width=170 mm]{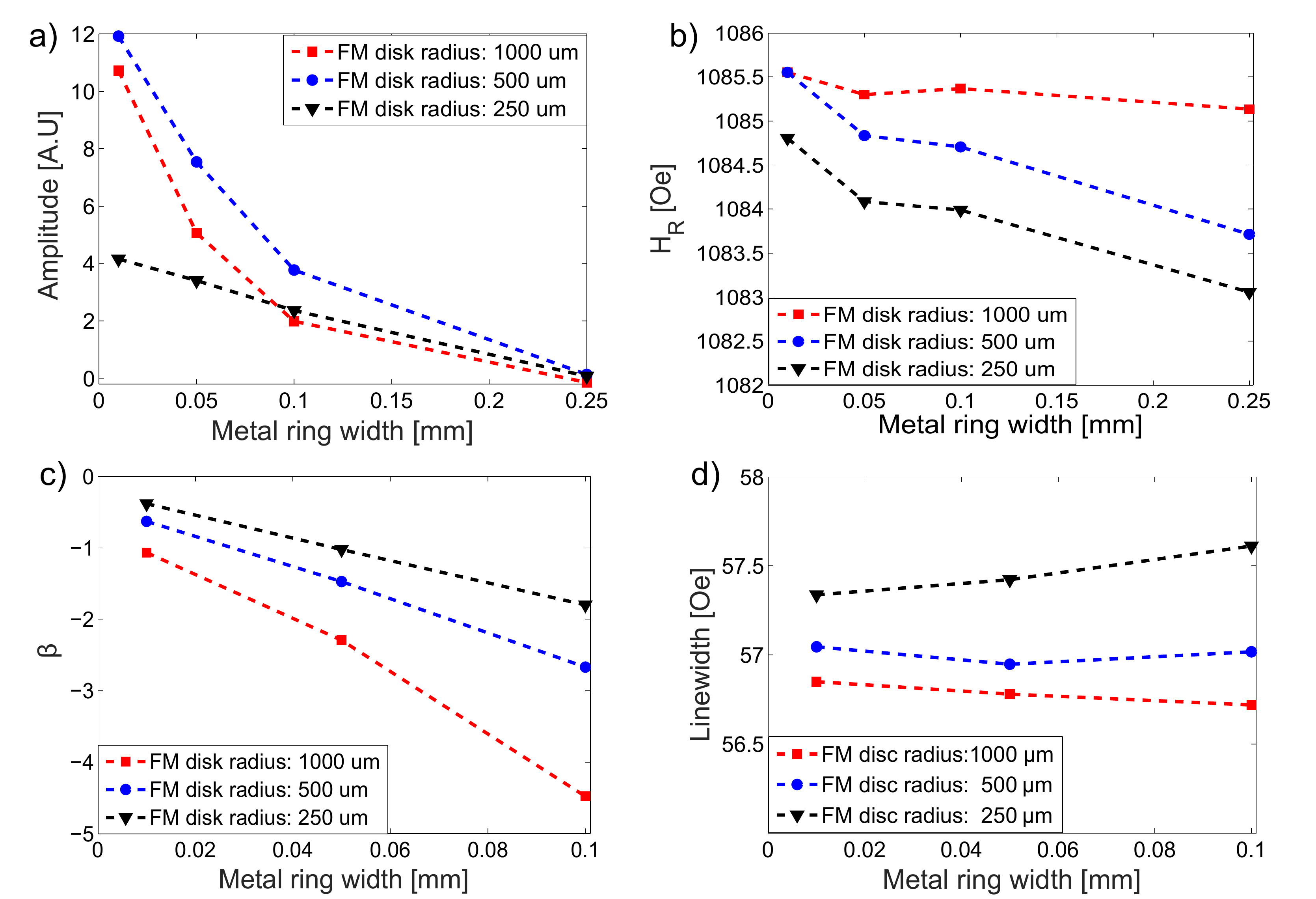}
\caption{\footnotesize a) FMR absorption amplitude, b) resonance field, c) lineshape asymmetry, $\beta$ and d) linewidth as a function of FM disc radius and metal ring width. }
\label{fig:halo1}
\end{figure*}

In Fig. \ref{fig:halo1}a we plot the FMR absorption amplitude for several samples of different size as a function of NM ring width, ranging from 10-250 $\mu$m. A clear trend is observed where the amplitude drops as the width of the NM ring is increased. For a NM ring of 250 $\mu$m, where the eddy currents should be strongest, the amplitude approaches zero.
We interpret this as an almost complete compensation of the applied MW field by the induced fields from eddy currents in the NM ring, suggesting a strong screening effect.

For the resonance field (Fig. \ref{fig:halo1}b) we observe an increase of $\approx $70 Oe compared to that of the previous sample series, which indicates a reduction in the saturation magnetization. The preparation of this sample series involves an extra lithography step, as mentioned in sec. \ref{sec:sampleprep}, and this could cause a reduced film quality compared to the other samples. A reduced film quality is consistent with the increased linewidth (Fig. \ref{fig:halo1}d), which is in the order of 10 Oe larger than for the other samples. 
However, we do not observe any clear trend in the linewidth as a function of disc size or NM ring width. This corresponds well with what one would expect if the change in linewidth for the PyAu bilayers in Fig. \ref{fig:PyAu_all}d is caused by the excitation of spin waves. In this last series, due to the sample geometry shown in Fig. \ref{fig:field_geometry_halo}, there will be no in-plane field components. Without any mixing of in-plane vs. out-of-plane field components, one should thus not expect any spin wave excitations and corresponding FMR linewidth broadening.
  
By investigating the lineshape assymetry in Fig. \ref{fig:halo1}c for a NM ring width ranging from 10-100 $\mu$m, we observe that the asymmetry increases with both the NM ring width and FM disc radius. (For a NM ring width of 250 $\mu$m, the absorption amplitude was to low to get a good estimate of the linewidth and lineshape asymmetry). As mentioned previously in sec. \ref{sec:PyAu} the sign of $\beta$ depends on the sample geometry, and we note here that $\beta$ has a negative sign, same as for the single Py discs in sec. \ref{sec:Py_single}. 
This is due to the similar sample geometry where the current loop (flowing mainly in the NM ring) is in the same plane as the FM disc. This is in contrast to the PyAu bilayers in sec. \ref{sec:PyAu} where the current is flowing mainly in the adjacent Au layer, and a positive $\beta$ is observed.  

The most interesting observation for this sample series, is the significant screening effect. The FMR absorption amplitude in Fig. \ref{fig:halo1}a indicates that the induced fields from eddy currents in the NM ring are able to almost completely compensate the applied MW field for a NM ring width of 250 $\mu$m, where the absorption amplitude drops to near zero. 
To estimate the compensation of the applied MW field by induced fields from eddy currents, we consider a simple current loop model:
The induced electromotive force (EMF) $\epsilon$ is given by the rate of change of the magnetic flux through the area enclosed by the current loop; its absolute value is given by $|\epsilon|=\pi r^2 |\frac{\partial h_{mw}}{\partial t}|$. The induced current is then given by, $I_{\text{Ind}}=\epsilon/R$ where R is the resistance of the current loop. The resistance of the Au ring is given by $R=2 \pi r \rho_{Au} /w \delta$, where $\rho_{Au}$ is the resistivity of Au, $w$ is the width of the NM ring and $\delta$ is the thickness of the Au layer. 
As discussed previously in section \ref{sec:PyAu}, the expressions for the magnetic field at the center of a circular current loop of radius $r$ in the x-y plane given by Eqs.(\ref{eq:Bz}) and (\ref{eq:Br}) simplify to the well know expression: $h_{\text{Ind}} = \mu_0 I_{\text{Ind}}/2 \pi r$. 
We are here interested in the condition where the induced field is comparable in strength to the applied MW field, $h_{\text{Ind}}\approx h_{\text{mw}}$. 
Fulfilling this relation, one obtains the following expression for the NM ring width: 

\begin{equation}\label{NM_width}
w=\frac{2 \rho_{Au}}{\mu_0 f_{\text{mw}} \delta}.
\end{equation}

A microwave frequency $f_{\text{mw}}= 9.4$ GHz, Au thickness of $\delta=10$ nm and the standard textbook value of resistivity, $\rho_{Au}$, results in a NM ring width of $w \approx 400$ $\mu$m.  Comparing the model calculation to experimental data, we observe such a compensation already at $w \approx 250$ $\mu$m. However, in this simplified model we only consider induced fields from currents in the NM ring. As we have discussed previously in section \ref{sec:Py_single}, and shown in Fig \ref{fig:amplitude2}a, induced currents in the Py disc also affects the absorption amplitude. Taking the compensation effect from the Py disc into account would result in that a thinner NM ring width is needed to compensate the applied MW field. This is consistent with the results from our simplified current loop model, which overestimate the NM ring width compared to the experimental result.

\section{Summary}
To summarize, we have shown that eddy-current effects can have a significant impact on the FMR excitation even in very thin metallic FM ($\sim 10$ nm Py), determined by the sample and MW field geometry.
Our results indicate how patterned FM/NM structures can be used to control the induced current paths. The corresponding MW fields have a relative phase shift with respect to the applied MW field, which depends on the sample geometry.
The induced fields produced by eddy currents can thus be used to compensate the applied MW field, as an effective screening of the FM in selected parts of the sample. 
We also provide evidence that controlling the local MW field enables the excitation of spin wave modes with a non-zero wave vector ($k\neq 0$), in contrast to the the uniform ($k=0$) mode normally excited in FMR experiments. 

What is not contained within this study is a proper investigation on the linearity of these effects. 
The induced current density  given by Eq.( \ref{eq:CurrentDensity}) depends linearly on the microwave  magnetic field. 
At some point however, the magnitude of induced currents will be limited by the conductivity and thickness of the NM film. This will limit the MW fields produced by eddy currents, and will thus introduce a threshold for when eddy-current effects will affect the system. 

Developing a more detailed theoretical model involving a numerical solution of the coupled Maxwells and LLG equation would also be beneficial. A numerical model would enable the investigation of the interplay between applied and induced MW fields for various sample geometries. This is however outside the scope of our current work.

For applications, combining the effects of eddy currents could be used for canceling of MW fields in different geometries, and providing highly localized fields due to the localization of the eddy currents. 
This allows for generating and controlling the MW field in small regions. In more complicated geometries, as often found in devices, the findings points towards that care should be taken in design, and that eddy-current effects can yield both FMR lineshape changes, unwanted or wanted screening/amplification of the local MW field, as well as the FMR excitation of spin waves with non-zero wave vectors. 
However, rather than treating eddy currents as a parasitic effect, our results suggest the possibility of actively using eddy currents to control the MW field excitation in thin film structures, which could be of importance for magnonics applications.

\section*{Acknowledgements}
This work was supported by the Norwegian Research Council (NFR), project number 216700. 
V.F acknowledge long term support from NorFab Norway, and partial funding obtained from the Norwegian PhD Network on Nanotechnology for Microsystems, which is sponsored by the Research Council of Norway, Division for Science, under contract no. 221860/F40.

\end{document}